\documentclass[10pt,letterpaper]{article}
\usepackage[top=0.85in,left=2.75in,footskip=0.75in,marginparwidth=2in]{geometry}

\usepackage[utf8]{inputenc}

\usepackage{cite}

\usepackage{nameref,hyperref}

\usepackage{microtype}
\DisableLigatures[f]{encoding = *, family = * }

\raggedright
\usepackage{changepage}

\usepackage[aboveskip=1pt,labelfont=bf,labelsep=period,singlelinecheck=off]{caption}

\makeatletter
\renewcommand{\@biblabel}[1]{\quad#1.}
\makeatother

\usepackage{lastpage,fancyhdr,graphicx}
\usepackage{epstopdf}
\fancyheadoffset[L]{2.25in}
\fancyfootoffset[L]{2.25in}

\usepackage[table,xcdraw]{xcolor}
\usepackage{color}

\definecolor{Gray}{gray}{.15}

\usepackage{graphicx}

\usepackage{sidecap}

\usepackage{wrapfig}
\usepackage[pscoord]{eso-pic}
\usepackage[fulladjust]{marginnote}
\reversemarginpar

\usepackage{float}

\begin{document}
\vspace*{0.35in}

% title goes here:
\begin{flushleft}
{\Large
\textbf\newline{Cancellation of principal in banking: Four radical ideas emerge from deep examination of double entry bookkeeping in banking.}
}
\newline
% authors go here:
\\
Brian P. Hanley\textsuperscript{1*}
%Author 2\textsuperscript{2},
%Author 3\textsuperscript{1},
%Author 4\textsuperscript{1},
%Author 5\textsuperscript{2},
%Author 6\textsuperscript{2},
%Author 7\textsuperscript{1,*}
\\
\bigskip
\bf{1} Butterfly Sciences
%\\
%\bf{2} Affiliation B
\\
\bigskip
* brian.hanley@bf-sci.com

\end{flushleft}

\section*{Abstract}
Four radical ideas are presented. First, that the rationale for cancellation of principal can be modified in modern banking. Second, that non-cancellation of loan principal upon payment may cure an old problem of maintenance of positive equity in the non-governmental sector. Third and fourth, that crediting this money to local/state government, and crediting to at-risk loans that create new utility value, creates an additional virtuous monetary circuit that ties finances of government directly to commercial activity. 

Taking these steps can cure a problem I have identified with modern monetary theory, which is that breaking the monetary circuit of taxation in the minds of politicians will free them from centuries of restraint, optimizing their opportunities for implementing tyranny. It maintains and strengthens the current circuit, creating a new, more direct monetary circuit that in some respects combats inequality. 
\\
\bigskip
Keywords: Double entry bookkeeping, loan principal cancellation, Modern Monetary Theory (MMT), taxation, taxes, tyranny. 
% now start line numbers
%\linenumbers

% the * after section prevents numbering
\section{Introduction}
This paper presents four conceptual ideas, and it is quite difficult to present one. These ideas lead to each other, and I believe that together they represent a way out of certain problems present in the modern world. These are inspired within the context of modern monetary theory (MMT), however, they also offer a practical alternative to MMT’s concept of placing all responsibility on the federal government to manage spending well in order to produce the desired positive equity in the non-governmental sphere. 

\marginpar{ 
\vspace{.5cm} 
\color{Gray} 
\textbf{First concept ---} Cancellation of principal. 
}

\emph{First}, is that the justification usually provided for cancelling principal of a loan does not hold up to close scrutiny. There are three reasons identified for its origin: A. Ancient history of debt as an obligation for repayment in goods or services, that is cancelled when paid. B. Fear of discovery in the murky history of banking. C. That the money created by a loan remains with the borrower that pays it off. On examination, none of these explanations is wholly convincing for banking. 

\marginpar{ 
\vspace{.5cm} 
\color{Gray} 
\textbf{Second concept --- }Principal can aid keeping the private sector in positive equity.
}
\emph{Second}, if cancellation of principal is revised, this money can aid curing the problem that the private sector needs to be in positive, not negative equity. MMT has recognized this, and has recognized that in order to keep the private sector in positive equity, then government must be in negative equity  \cite{Keen2020ModelOfModernMonetaryOperations}. However, MMT management of money through deficits providing private sector positive equity faces multiple political obstacles, and it creates a very serious potential for the rise of tyranny when the national government fiat currency issuer is no longer motivated by taxation needs to keep the economy operating for the good of the people \cite{Hanley2021IsModernMonetaryTheorysprescriptiontospendwithoutreferencetotaxreceiptsaninvitationtotyranny}. 

\marginpar{ 
\vspace{.5cm} 
\color{Gray} 
\textbf{Third concept ---} Principal payments could fund local and state governments directly. 
}
\emph{Third}, when revised, as principal from loans is paid down, this money can be credited to funding the operations of local and state government below the federal level. This creates a built-in monetary circuit that directly motivates governments at the city, county, and state level to ensure that business loan activity and loan service are optimized for the society. This can also allow modern government to keep its current beliefs about the how the circuit of income at the national level works, and maintain positive equity. 

\marginpar{ 
\vspace{.5cm} 
\color{Gray} 
\textbf{Fourth concept ---} A new type of bank that does not use the central bank's discount window. 
}
\emph{Fourth}, create a new category of bank, which does not have access to the central bank’s discount window, only using the central bank for settlement. Such a new bank would be strictly chartered for investment into new business enterprises that create new utility value in the economy by making at-risk equity loans to said enterprises. This new type of bank could do two operations not allowed in normal banking. It could write off it's own bad loans without repayment, provided no part of it was a loan to another party, and it could retain full value of loan principal when repaid by its equity, or from money derived from this equity. 

Here I will discuss these concepts, and delve into the operations of the Medici’s to ground the discussion in historical precedents. 
\subsection{Roots of banking’s bookkeeping rules.}
The deep history of double entry bookkeeping in banking is cloudy. McCleay is a good reference \cite{McLeay2014MoneyCreationInTheModernEconomy}, however, examining the history and development of banking, the neat and clean version of what civilization does today does not entirely agree even with our modern rules. This should not surprise those with historical knowledge of finance and banking, however, historical “excursions” are generally viewed strictly as aberrations or even crimes in our time. I prefer to look at them with the kind of fresh eyes that Tooke brought to his analysis of the Bank of England \cite{Tooke1844InquiryIntoTheCurrencyPrinciple}. 

Why is cancellation of loan principal done? Ask the question and various answers will be given, ranging from, “it’s obvious” to apparently more erudite responses. Double entry bookkeeping has a self-consistent elegance that once grasped tends to cause questions to drift away. 

 Once created by a bank loan, the money does not disappear unless made to do so by accounting. This is so intuitively obvious that a number of economists have argued that the money does not disappear when paid back. And yet, even for those that accept this paradigm, there are things that can’t be made to disappear which can be received as payment for principal, such as gold, silver, or any concrete physical asset. To make it cancel, the asset must be sold and then the money received for it can disappear. 
 
\subsection{Why cancel loan principal upon payment?}
The best explanation for this came from Geoffrey Gardiner, who said that the customer who borrows is the primary creator of money, the bank being a sort of midwife \cite{Gardiner2019EdgeMidwife}. So the money created (principal) \underline{goes to the borrower}, or some downstream connection of the borrower, which when paid back requires the bank to cancel the principal. This logic seemed questionable to me, and we discussed that if the borrower is able to pay back the principal, that could also be construed to mean that the principal creation was validated. 

Accepting the above explanation, if we predicate that uncle Fred in St. Louis borrows \$85,000 from Bargely’s Bank, with a business plan to start Popsicles 21, as he pays back the principal and interest, the principal slowly gets paid down to zero. The bank retains the interest. Where the interest comes from is an interesting question not treated here.

One could predicate that if a non-bank entity, such as Mary in San Francisco, loans \$85,000 of her own money to uncle Fred to start Popsicles-21, then when uncle Fred pays Mary back, she should cancel that \$85,000 that is her principal. However, Mary would then be left with nothing but the interest, which is obviously not going to work for her. If we implement this rule universally, it would mean that any non-bank making a loan is as good as burning money. 

Consequently, it is obvious that the reason for cancellation of principal has something to do with the origination of the funds by a bank, which has created new money by the figurative stroke of a pen. However, perhaps it is even older and more basic than that. 

Graeber includes in his writings on the origins of money as debt instruments, that debts are seen in their earliest form as tribal accounting for obligations, brides, animals and other items \cite{Graeber2014DebtTheFirst5000Years}. When such debts are paid back, this cancels the debt, thereby setting the ‘principal’ of the obligation to zero. This appears to be a possible root of the practice of cancellation of principal in banking. 

So let us look into the history of banking at one of the families that is sometimes credited with creation of modern banking. 

\subsection{The beginning, the Medici, and banking}
Sometimes the Medici’s are credited highly in banking, but they aren’t known to have invented anything for certain. However, I suspect they ‘invented’ some part of loan profitability that we no longer recognize. The Medici made good use of what they knew and got famous, probably because of their lavish spending on architecture, the arts, and their meteoric rise to wealth and power. 

As far as I can determine, there is not any specific place where “banking” begins and vault storage money-changing with its bills of exchange is subsumed/replaced. Instead, it seems to be an evolution that for the Medici (and the rest of Italy’s bankers) was enabled by the presence of Rome nearby which received tithes/tribute from all of Christendom.  This created a built-in cash flow which then had to be spent, and filtered back out. 

The beginning of banking isn’t even in Rome \cite{TEMIN_2004FinancialIntermediationEarlyRomanEmpire} and probably not in Mesopotamia \cite{Naik2014BeginningBankingMesopotamia} either. The earliest possible forms that may be tally-stick tokens date from the paleolithic \cite{deHeinzelin2014Ishango,Beaumont1973BorderCaveIshangoLebomboBones} and down into the near current era \cite{Baxter1989TallyAndCheckerboard,Jenkinson_1911ExchequerTallies}. 

Part of the problem is also to define what a bank is. In its most basic form a bank (in current era terms) is something that makes loans, generally invents/creates money to do so, and has the confidence of its society. This arguably includes things like tally-sticks into the category of accounting for such creation of funds for trade. 

As for how banking works internally in any era, to a great extent I think we have to say that a bank can do whatever it is allowed to do by custom and/or regulation --- or else whatever it manages to get away with. Even today the formal rules are not always enforced strictly, with the 2008 global financial crisis (GFC) being a case of repeated non-enforcement of regulations \cite{Hanley2012ReleaseOfTheKraken}. In the Medici’s era, there was no enforcement except for the collapse of a bank\footnote{Gentile di Baldassarre Boni died in a debtor’s prison after leaving the Medici partnership.}. Nobody was auditing the Medici books, and the modern practice of having one set of books for taxes and another for the business has ancient roots. In the Medici’s case, it’s obvious that when someone writes in the margin of an account book, “For love of the tax man,” regarding some particular bit of flim-flam, the tax man was not supposed to see that. So we know the Medici kept multiple books.

Call it ‘regulatory capture’ or call it ‘the real world’, banks do things they aren’t technically supposed to do. Few people have memorized all the rules of the Federal Reserve’s manual of over 1000 pages. Regulators can look forward to a more lucrative position on the other side of the table if they play their cards right. Sometimes this is flagrantly obvious, as in the non-prosecution of anyone involved in creating the 2008 GFC. And what banks are supposed to do is not always entirely clear. Things get invented that create conditions not seen before, as in the runup to the GFC \cite{Hanley2012ReleaseOfTheKraken}. 

Where bankers end and government begins is often a fuzzy boundary. This is decried by some, but it is also how our modern world was created, and without it, it is questionable if the modern world could exist. Again this goes back to before the Medici. Romans, Mesopotamians, and probably the makers of Paleolithic tally-sticks gained power through finance (appendix \ref{Sect:Virtual_currency}). The Medici are perhaps an archetypal example of this – making themselves a power in Italy, and across Europe. 

\section{The Medici data}
Initially, I presumed that the Medici banking business was seasonal, and this idea makes sense. But the data says that this effect is mild. Within Christendom the payment of tithes to Rome created a cash flow source in Rome, and the Medici were part of making it flow back out, thus helping to enable more tithes/tribute to be paid. The Medici capitalized on flows of hard money stocks \cite{Parks2014MediciMoney,DeRoover1966RiseDeclineMedicidBank}. 

In that sense the Vatican in tandem with the bankers of Italy played the role of the “great circulator” of large amounts of money. I think that we can safely say that the practice of tithing forced money velocity to a level that was considerably higher than it otherwise would have been, and Italy was a major beneficiary of this. Temples had done this for a long time going back to the Hebrews. 

The Medici also practiced secret bank accounts as banking havens do today. Bishops and even the Popes hid money this way. The condottieri (mercenary armies of the time) turned their money over to the bankers, as did many others. So money was safe from nobles or the Pope confiscating it, as was often done with real property. In some sense these accounts were the precursors to our savings accounts. Although it was at the discretion of the banker to give them interest or not. Here, to be conservative, it is assumed that interest is paid on depositor money when  used as reserves for making loans. Whether this was done in the Medici period when the agio \ref{Sect:Agio_shubati} was a source of income is unknown.

Being in this pipeline of gold and silver flowing into Rome is a primary way that the Medici were able to keep handing out gold (Florins) and silver (piccioli) when the international flow of gold and silver in trade was, net, away from Florence to the manufacturers and raw materials suppliers of the time in the rest of Europe, or into the surrounding countryside. (This included payments to mercenaries to fight endless internecine wars.) They also performed triangle trades, going into business to use their loan capacity to generate cash flow and profits. 

These early bankers got around usury laws (and religious excommunication) by giving exchange notes and making de facto loans that were usually fairly short term (3-6 months). Formally those loans were written as currency exchanges set up to profit above the real rate of exchange. When there was not a real exchange of currency, they performed “dry exchange” (cambio secco), in which fictional currency exchange transactions took place. Net, this acted like interest. So they implemented effective interest rates, but camouflaged. Their loans appear to be short-term, fixed price loans. Gardiner guessed the Medici would roll loans over, however, nothing suggested this in the data. Much, if not virtually all, of these transactions were secret. Depositors of gold did so secretly, and loans were also given secretly.  This latter is still mostly the case, we call it privacy. The difference today is that auditors can look into everything, and records can be demanded by authorities.  

\subsection{How much interest did the Medici get on their loans?}

\begin{figure}[H]
\includegraphics[width=130mm]{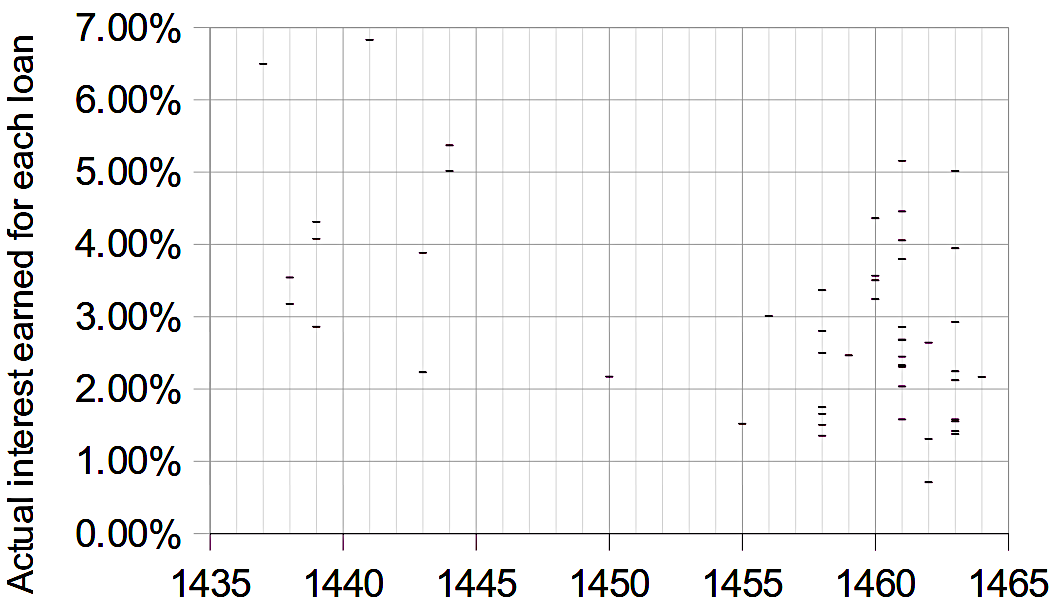}
\captionsetup{labelformat=empty} % makes sure dummy caption is blank
\caption{} % add dummy caption 
\label{Fig_1_Actual_earnings_percentages} 
\end{figure}

Answering this question is complicated, because sources vary, and summaries do not agree with detail figures. Figure \ref{Fig_1_Actual_earnings_percentages} data is drawn from de Roover’s exchange tables for Venice/London/Bruges and a diagram of a dry exchange transaction \cite[(p117-120, 133)]{DeRoover1966RiseDeclineMedicidBank}.	 
\marginpar{ 
\vspace{.7cm} 
\color{Gray} 
\textbf{Figure \ref{Fig_1_Actual_earnings_percentages}. Actual earnings.} 
Actual earnings percentages for specific loans by year. Calculated as nominal loan rate per day $\times$ duration of the loan.  Simple interest, no daily compound.
}
 % avoid blank space here

The work of de Roover is more accurate than Parks, because Parks used de Roover as his source, and appears to have misinterpreted some material.  I’ll start with a set of graphs to give a sense for what loan rates and durations were. 

\begin{figure}[H]
\includegraphics[width=130mm]{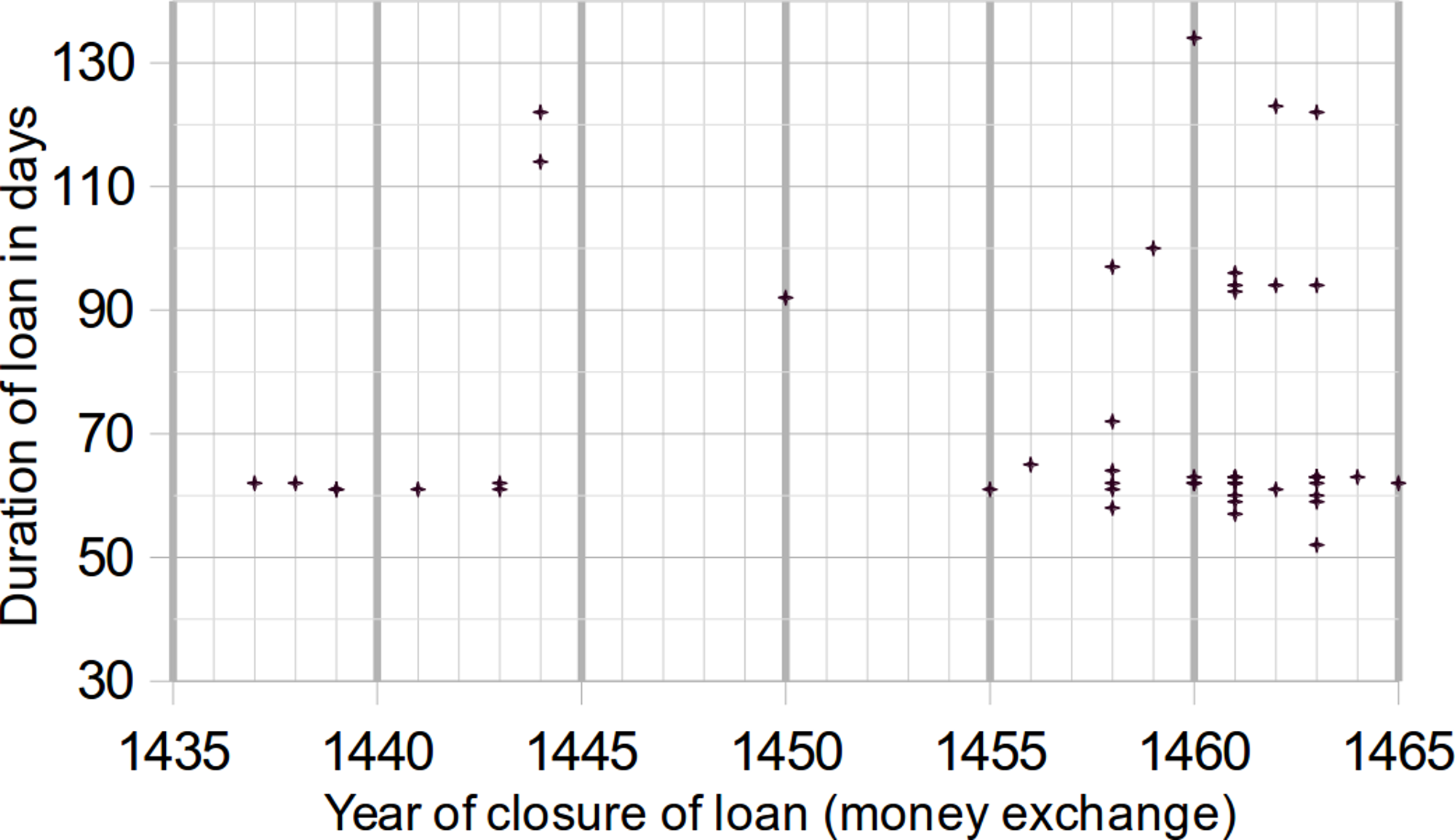}
\captionsetup{labelformat=empty} 
\caption{} 
\label{Fig_2_Year_of_loan_closure} 
\end{figure}

\marginpar{ 
\vspace{.7cm} 
\color{Gray} 
\textbf{Figure \ref{Fig_2_Year_of_loan_closure}.Year of loan closure versus duration.} 
}
 % avoid blank space here
In figure \ref{Fig_2_Year_of_loan_closure}, there appear to be 3  common durations for these loans. ~ 60 days ($n=39$), ~95 days ($n=9$), and ~ 120 ($n=5$) days. With a total of 53 loans, 74\% are 60 day range, 17\% are 95 day range, and 9\% are 120 day range. At 98\% confidence, this sample of 53 loans has a 16\% margin of error.  I will assume this 27 year period had 1000 voyages per year from Italy, or 27,000 voyages total to be financed. This 1000 voyages number is not sourced, but is required to model reported earnings below.

Figure 3 shows the distribution of nominal yearly loan rates. However, it is not realistic to believe that the Medici bank could loan their money all year round at the rates shown in figure 3. The Medici loaned money primarily for trade voyages. 

\begin{wrapfigure}[14]{l}{90mm}
\includegraphics[width=90mm]{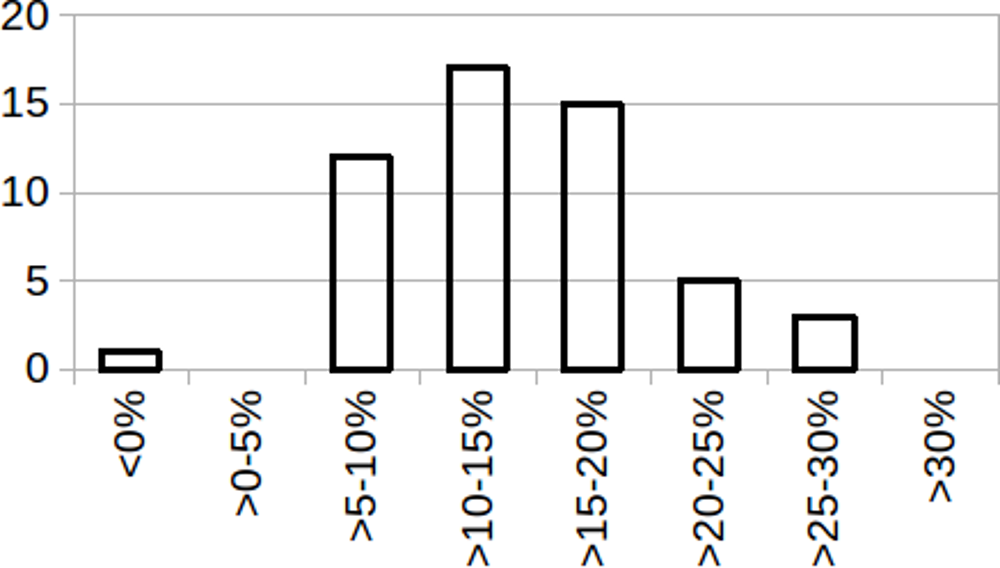}
\captionsetup{labelformat=empty} 
\caption{} 
\label{Fig_3_Interest rate histogram} 
\end{wrapfigure}

\marginpar{ 
\vspace{.7cm} 
\color{Gray} 
\textbf{Figure \ref{Fig_3_Interest rate histogram}. Interest rate histogram projected to a full year.} 
}
 % avoid blank space here

Initially I had the impression loan activity was quite seasonal. But what is seen in Table 1 is that while Spring is a bit of a lull, we should see a good number of loans outstanding for this period. 

\begin{wrapfigure}[15]{l}{100mm}
\includegraphics[width=100mm]{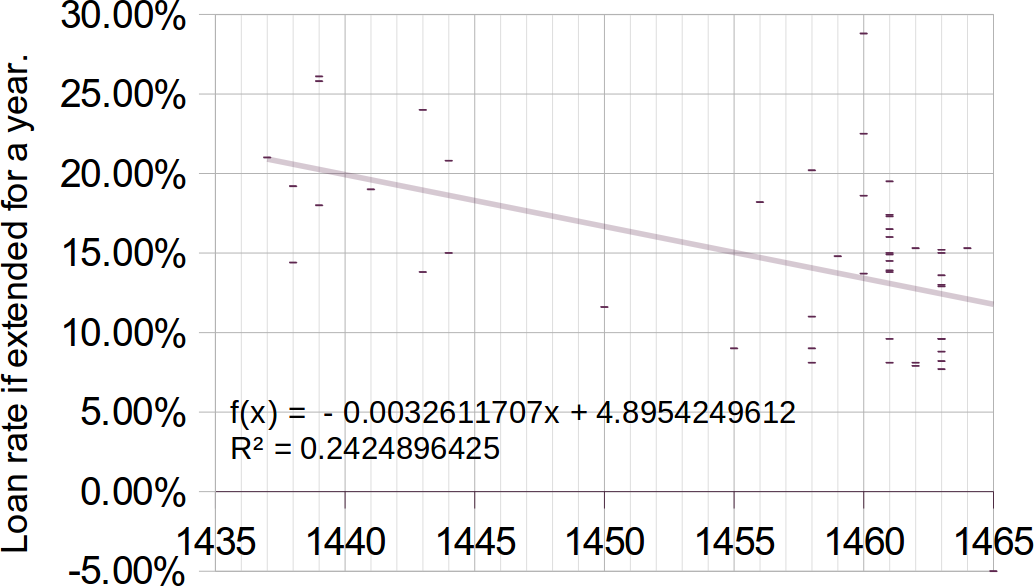}
\captionsetup{labelformat=empty} % makes sure dummy caption is blank
\caption{} % add dummy caption 
\label{Fig_4_Loan_rate_extended_to_full_year} 
\end{wrapfigure}

In figure \ref{Fig_4_Loan_rate_extended_to_full_year} is shown the loan earning percentages corrected for duration. (i.e. Figure \ref{Fig_4_Loan_rate_extended_to_full_year} shows the nominal yearly loan rates by year.) There may have been a downward trend during this period of the Medici wealth accumulation. 
\\
\marginpar{ 
\vspace{.7cm} 
\color{Gray} 
\textbf{Table \ref{tab1}. Seasonality of loans, number of loans.} Columns contain counts of loans with a start date or end date in the season designated.
}
\begin{table}[H]
\caption{}
\begin{tabular}{c|c c}
       & End date & Start date \\ \hline
Winter & 18       & 15         \\
Spring & 9        & 11         \\
Summer & 12       & 13         \\
Fall   & 14       & 14         \\ \hline
Total  & 53       & 53        
\end{tabular}
\label{tab1}
\end{table}

Figure \ref{Fig_4_Loan_rate_extended_to_full_year} was calculated using the data of figure \ref{Fig_1_Actual_earnings_percentages} and table \ref{tab1}. This figure \ref{Fig_4_Loan_rate_extended_to_full_year} graph reflects more accurately what the real earnings could have been for the Medici bank.  This data comprises the 53 loans for which we have records. (The Medici made money as traders as well, but it’s what is available.)

\marginpar{ 
\vspace{.7cm} 
\color{Gray} 
\textbf{Figure \ref{Fig_4_Loan_rate_extended_to_full_year}. Actual earnings.} 
Actual earnings percentages for specific loans by year. Calculated as nominal loan rate per day $\times$ duration of the loan.  Simple interest, no daily compound.
}
 % avoid blank space here

The Medici’s are reported to have made loans to royalty \cite{Medici1999ThoseMediciEconomist}, however, de Roover says they went into this risky business only after the Florentine galley trade failed. He thinks the galley trade likely failed because of a shortage of money in the low countries \cite[p150-151]{DeRoover1966RiseDeclineMedicidBank}. Florence’s failure of trade occurred circa 1775 or so, which is later than any detail data available on the bank found in de Roover.  So for the period we see here, the Medici were also traders.

Parks claims much higher loan rates than I see in de Roover. Parks says there were fixed-price loans returning between 7.7\% and 28.8\% interest on a single transaction. This range from Parks is de Roover’s projection of what the yearly rate would be. Parks appears to have mistakenly used those projections as the value for individual loans over a few months time.  This is confirmed by figure \ref{Fig_4_Loan_rate_extended_to_full_year} data. 

We know that the Medici became highly influential in government. We know they made money as merchants. And most of their loans were written as notes, not least because it is much harder to use a stolen note. 

However, those figures do not account for the inevitable losses. Ships would be lost at sea, although according to de Roover, this was rare. He logs one incident of piracy by a corsair, and cargoes were insured. In addition, people would die of plague. The century of the Medici was the century after the Black Death, which hit Italy first. With half the population of cities dying in the previous century, lenders had to experience significant losses that harmed their profits, although this could be counteracted by freeing up of real estate and seizing property. Additionally, when a person with a secret account died, did bankers of the era diligently attempt to find heirs? This could have been a secret source of income for bankers of the Medici period. 

\subsection{What rates did the Medici need to make the profits documented?} 
We can only estimate, because data is incomplete \footnote{De Roover’s accounting of the Medici is incomplete because the Medici records are incomplete. I suspect digging into this would require at least a season of study of museum pieces, if that could be arranged. And given that others haven’t found them, and my lack of ability to read 600 year old Italian script, I don’t know that it could be productive.}, and their earnings are mixed up with their profits from trade. So, these figures must be understood to treat all of the Medici earnings as if they were loans. The starting point figures are in Table 1. I will develop this step by step to attempt to show the range of possible results for the Medici bank. 
 
\subsubsection{Model 1 – simple interest compounded yearly.}
Initially, Table \ref{tab_Med_bank_by_year} looks straightforward. We start at near 10\% per year, and gradually increase to approximately 27\% per annum for the period from 1435-1450. Not impossible to believe for a family that arguably took over Florence. 
 
\begin{table}[H]
\begin{adjustwidth}{-1in}{0in}
\caption{\textbf{Medici bank earnings, by year.} Profits multiple: $\frac{Capital + Earnings}{Capital}$ Italic items are from cited sources. Everything else is calculated.}
\begin{tabular}{lllllll}
 &              &  &  & Profits  &  &                           \\
 &              &  &  & multiple &  &                          \\
 &               & Time & &  since  & Earnings &                           \\
 & Capital               & elapsed & Earnings in & previous  & per &                           \\
Year & florins               & Months & florins & notice & annum & Comment                          \\ \hline
\emph{1397} & \emph{8000} & 12                  &                     &                                        &                    & Bank founded                       \\
\emph{1398} &                              & 6                   & \emph{1200}                & 1.15                                   & 09.76\%   &                                    \\
\emph{1420} & 8000                         & 276                 & \emph{152820}              & 20.1025                                & 13.94\%   & Giovanni retires, Lorenzo inherits \\
\emph{1435} & 8000                         & 180                 & \emph{186382}              & 24.29775                               & 23.70\%   &                                    \\
\emph{1450} &                              & 180                 & \emph{290791}              & 37.348875                              & 27.30\%   &                                   
\end{tabular}
\label{tab_Med_bank_by_year}
\end{adjustwidth}
\end{table}

But, there’s a big problem \footnote{The Medici’s bank was started with 10,000 florins. However, Gentile (Gen) pulled out soon, leaving \$8000 florins of capital. Investor contributions: Gio: +5500, Ben: +2000, Gen: +2500, Gen -2500, Gio: +500}. These figures in the earnings column are total profits distributed for the previous period. And in the Table \ref{tab_Med_bank_by_year} calculations, no money is taken out as earnings until the 13 or 15 year time span is over. Nor do we know how much was left in the bank as capital, nor how much was on deposit. 

What becomes clear on reflection is that Table \ref{tab_Med_bank_by_year} can’t be right because the Medici removed money every year in profits. So we have to calculate it, which means we have to ask the question of how much the Medici took as their profits? 

In de Roover, the tax record for Lorenzo in 1457 indicate he took F2800 for that year. This 1457 income number submitted to the taxing authorities is quite far from the necessary average distribution from the bank.  If we assume the reported earnings of F290,791 for the period 1435-1450 was divided equally among the 15 years in the period, this would be F19,386 per year, which is 6.9 times the reported income. There were multiple partners, and each would take dividends based on their share ownership. If we assume reported income was half of actual income, 5 total partners, Cosimo and Lorenzo taking the same amount, and the other 3 taking half of what Lorenzo and Cosimo received, the total yearly payments to partners would be essentially all of earnings. We are left to guess, as de Roover, Parks, and others indicate tax evasion in that period was normal. Below I choose a low of 2.5\%, and a high of 10\%, while the S\&P 500 retention rate is 4\%. 
	
We know that massive spending occurred under Lorenzo and Cosimo. They finished the Duomo. They doubled the size of the Pitti Palace. The Medici commissioned art that still impresses us today. Their parties were legendary. The Medici were a public face of the Renaissance. Their lavish spending, and probably their money creation, doubtless had a significant effect on the Florentine economy by increasing both velocity and quantity of money in circulation.  

\subsubsection{Model 2 – Rates required based on some percentage retention of earnings.}
Two more year on year spreadsheet models, broken into 3 intervals of 13 or 15 years, using published data, were made using three assumptions of how much money the Medici’ might have retained each year, 2.5\%, 5\% and 10\%, for the purpose of growing their capital. This allowed calculation of what yearly interest rate was needed for each of the three intervals based on the Medici’s capital. This model assumes no lending based on deposits. (Yes, this also is not entirely realistic, but it’s the next logical step and will be developed further.) 

\marginpar{ 
\vspace{.7cm} 
\color{Gray} 
\textbf{Table \ref{tab3}. Interest \& retention.} Interest rates required to meet reported profits taken, varying percentage of bank capital retained each year.
}

\begin{table}[H]
\caption{}
\begin{tabular}{llll}
\textbf{Period} & \textbf{2.50\%}           & \textbf{5.00\%}           & \textbf{10.00\%}           \\ \hline
1397-1420       & 72.66\%                   & 63.90\%                   & 51.11\%                    \\
1420-1435       & 91.13\%                   & 64.85\%                   & 40.69\%                    \\
1435-1450       & 99.82\%                   & 63.09\%                   & 36.11\%                   
\end{tabular}
\label{tab3}
\end{table}
  
What can be seen in table \ref{tab3} is that a 5\% retention rate at least gives us a stable interest rate to deal with. However, all of these rates are far above both the 28.8\% interest projected for the most lucrative loan in the records, which is 28.5\%. They are also wildly beyond the actual per-deal loan earnings shown in figure \ref{Fig_4_Loan_rate_extended_to_full_year}. For now, we will just note this.
   
\subsubsection{Model 3 – Retention figures with fraction of depositor money.}
The next question is, how much depositor money (as gold and silver) did they have as a basis to lend? Honestly, this is not possible to tie down. In Florence, I was told the figure $\frac{3}{4}$ by an historian as an estimate of how much was depositor’s money. So, for each of the three capital retentions of table \ref{tab3}, I did a set of 3 models that assumed $\frac{1}{2}$, $\frac{3}{4}$, and $\frac{7}{8}$ of their loans were made from depositor’s money. I assumed that half of the interest received would be paid to depositors, and retentions would be taken out of Medici earnings based on the amount of Medici capital. The results are in table \ref{tab4}.

\marginpar{ 
\vspace{.7cm} 
\color{Gray} 
\textbf{Table \ref{tab4}. Interest, retention \& earnings.} Interest rates required to meet reported profits taken, varying percentage of capital retained per year, and varying amount of deposits relative to capital.
S\&P average retention rates on earnings are 4\%.}

\begin{table}[H]
\caption{}
\begin{tabular}{llll}
                & \multicolumn{3}{l}{\textbf{2.5\% of capital retained/yr}}                                       \\
                & \textbf{Deposits =} & \textbf{Deposits =} & \textbf{Deposits =} \\
\textbf{Period} & \textbf{capital} & \textbf{3 x capital} & \textbf{7$\times$ capital} \\ \hline
1397-1420       & 45.62\%                     & 27.19\%                         & 12.35\%                         \\
1420-1435       & 53.04\%                     & 31.61\%                         & 14.35\%                         \\
1435-1450       & 55.72\%                     & 33.21\%                         & 15.08\%                         \\
                & \multicolumn{3}{l}{\textbf{5\% of capital retained/yr}}                                         \\
                & \textbf{Deposits =} & \textbf{Deposits =} & \textbf{Deposits =} \\
\textbf{Period} & \textbf{capital} & \textbf{3 x capital} & \textbf{7$\times$ capital} \\ \hline
1397-1420       & 46.25\%                     & 27.37\%                         & 12.42\%                         \\
1420-1435       & 53.50\%                     & 31.66\%                         & 14.37\%                         \\
1435-1450       & 55.72\%                     & 32.98\%                         & 14.96\%                         \\
                & \multicolumn{3}{l}{\textbf{10\% of capital retained/yr}}                                        \\
                & \textbf{Deposits =} & \textbf{Deposits =} & \textbf{Deposits =} \\
\textbf{Period} & \textbf{capital} & \textbf{3 x capital} & \textbf{7$\times$ capital} \\ \hline
1397-1420       & 39.68\%                     & 23.14\%                         & 10.48\%                         \\
1420-1435       & 36.29\%                     & 21.17\%                         & 9.59\%                          \\
1435-1450       & 33.79\%                     & 19.71\%                         & 8.93\%                         
\end{tabular}
\label{tab4}
\end{table}

The rates seen in table \ref{tab4} mostly exceed the maximum projected (quite unrealistic) yearly rate of 28.8\%, or come in not far below it. Given that the mean projected yearly rate was ~15\%, only the 7$\times$ capital column can be taken seriously. However, is it realistic to think that the Medici could get the table \ref{tab4} average earnings in a year when their business was composed of such short-term loans? The average loan of $2\frac{1}{2}$ months returned just 2.98\% on the transaction. So we must ask, what was their yearly utilization?  

By treating all of these loans as a single year, I derived a maximum loan utilization (Table 4) over the course of a year of 64.58\%, which rounds up to 65\%. To derive this, I mapped the months used for each loan, added up each column, and the maximum coincident loans was 16. The average yearly loan rate is then 15.07\%. 

Table \ref{tab5} shows the basis for attempting to estimate a maximum loan activity in a year by treating all of the Medici data as if it were a single year. Since the maximum number of “simultaneous” loans was 16, in January, I used this as the standard for what full utilization would be on the total Medici bank. Summing the fraction row and dividing by 12 months yields 64.58\% loan utilization for the year. The average nominal yearly interest rate is 15.07\%. 64.58\% of this average interest provides a 9.73\% yearly actual yield. With 10\% of earnings retained each year, this interest rate is barely enough to cover the 7$\times$ capital deposits column 4, for 2 of the 3 quindecums (15 year periods).

\begin{table}[!ht]
\begin{adjustwidth}{-1.0in}{0in} % comment out/remove adjustwidth environment if table fits in text column.
\centering
\caption{{\bf Treatment of all available Medici loan data as a single year.} This table assumes that 16 loans at one time are the maximum. It is assumed that all loans average to a similar amount, so they can be treated equally. This also assumes that 16 loans represents all of the Medici capacity for loaning money, however, this latter assumption is generous.}
\begin{tabular}{rrrrrrrrrrrrr}
                & Jan  & Feb  & Mar  & Apr  & May  & Jun  & Jul  & Aug  & Sep  & Oct  & Nov  & Dec  \\ \hline
Counts          & 16   & 13   & 11   & 5    & 5    & 10   & 10   & 12   & 12   & 8    & 9    & 13   \\
Fraction of max & 1.00 & 0.81 & 0.69 & 0.31 & 0.31 & 0.63 & 0.63 & 0.75 & 0.75 & 0.50 & 0.56 & 0.81
\end{tabular}
\label{tab5}
\end{adjustwidth}
\end{table}
	
 However, it also isn’t realistic to think that the Medici had the same fraction of deposits for the entire 53 years recorded. Any business tends to proceed in jumps, not smooth curves. And yet their earnings were consistently high.  

 One potential explanation is that all of these loans that were written as short term loans were just rolled over immediately, and loan utilization was near 100\% at all times. This was suggested as a possibility by Geoffrey Gardiner. High utilization is found in current banks, so it makes some sense to us to assume it. However, aside from there being no strong rationale for this idea, the data appears to contradict it. The Medici’s loaned short term for specific voyages, and sometimes for business ventures, at least some of which were their own. 
 
Even if the Medici had been able to keep their money earning interest all year round, the interest rates that we see in figure \ref{Fig_4_Loan_rate_extended_to_full_year} are much lower than the rate required to break even (100\% interest) in my hard-money fair game \cite{Hanley2015ZeroSumMonetarySystem}, which is another piece of evidence indicating that the Medici were, de facto, performing a kind of reserve based lending, which is what they are credited for. 

Perhaps they kept more capital in their bank than is shown. However, I think that the idea that they added 10\% of their earnings as capital each year is not likely. I base this on seeing what they built, and learning about the period and their free-spending ways. 

I think something else was going on in addition to their pioneering fractional reserve banking. I think the Medici discovered that they could keep much of the principal paid back from loans as “profit”. They could not keep all of the principal created by their loans, and I’ll explain why, but I think that a major factor that made them rich is keeping variable fractions of of the loan principal when it was paid back in gold or physical assets such as trade goods. 

\section{Development of banking – a fairy tale as others in economics?}
I’ll go through a thought exercise that I base on my studies.  In Florence, the Medici were said to believe they were running a scam and putting one over on the people of the city. They lived in terror of  being found out that they had more loans outstanding than they had money in their coffers. This sounds reasonable, but was that all they were worried about? Let’s also remember that the Medici were not historically a banking family. When they got into it, they would have the mind of outsiders, looking at everything with fresh and questioning eyes. Also, in the beginning, they would not have the influence that they developed in Cosimo or Lorenzo’s time. They bought their way into influence in government. 

I was told they never let their loan accounts get over double what they had in their vaults. And they enforced their contracts mercilessly. The stories of thuggish retribution vary from beatings of naked debtors with birch in the public square, perhaps chasing them through the streets, up to the horrible fate of being dropped from 10 meters height with hands and feet tied together and left to die in agony, or live for a time as a cripple in a world not kind to such. What their real methods were I cannot say for sure. It was a rough period in history \footnote{I suspect that the Medici’s Uffizi backing up on the river Arno may have been useful at times. Heave-ho, off you go, sort of thing for the overly difficult.}.  

With that as backdrop, what was it that these early bankers like the Medici’s were doing? Let me try to reconstruct it through a thought process. They began as vault operators that issued bills of exchange. In their time, powerful families of wealth would have vaults where they would store gold and silver for others. A second type of local bank would deal in silver exchange into gold, and also handle gems. A third type was the pawnbroker (lombard). These large vault operators were in communication with other vault operators (banks). 

A person could deposit gold in their vault or, for a fee, get a letter of credit. With that he went to another vault operator, or sent the letter to another merchant in a different city, and that vault operator would give him gold, again for a fee. It was a system that grew up to prevent robbers from getting the gold. Today, the Arab hawala system functions much the way that ancient vault operator bankers did.  

The Medici also made loans of other people’s money (see above discussion) and took a cut of the interest earned. They did this with paper and bookkeeping mostly. Plus there were checks, and printed paper currency notes redeemable for gold coins. Homes of that period were castles with battlements to keep out mercenaries, and everyone had to be on their guard. So their paper notes were popular. F1,000  weighed 3.5 kg. With F1,000, a man could have a palazzo built, a nice home of stone with battlements. So using accounts at the bank rather than carrying it around made sense when possible. 

\subsection{Making a loan and retaining Smaug’s hoard}\footnote{Smaug is a large dragon from "The Hobbit" by J.R.R. Tolkein that lived under a mountain and slept on his hoard of stolen gold.}
Let’s think about what a guy like Giovanni de Medici would observe when he made a paper loan. He would see a man go off with his note, and he made a bookkeeping entry for the borrower’s debt. He might give the man paper Florins for most of it, along with some metal coins, or just a letter of credit. And yet, in his vault sat most of what he loaned. The gold and silver backing the notes just sat there, like Smaug’s hoard. One wonders if the Medicis went in there to just contemplate their world. 

Then, Giovanni would get paid back, principal plus interest2. I was told in Florence that particularly in the early days the Medici liked to get paid back in gold and silver. Let’s diagram how that would work.  We will not use deposits here. We will assume that the borrowers go off with their paper to their homes or place of business. There are no regulators watching over him, so he can do his books however he wants to. Here, the banker gives the borrower a loan consisting of F10 in gold coin, and F90 in paper form. 

What table \ref{tab6} shows is that somewhere out there in the Italian countryside, or on a ship to Bruges, a merchant who borrowed F100.0 from Giovanni’s bank exchanged the F90 paper for gold by doing business. And he earned something to make it worth his while. Maybe he earned 25 florins, a good amount of money in that era. So let’s say Giovanni gave the family that had the particular F100 deposit of gold half the interest that he earned. That’s F0.19 that he credits to the depositor family. And he keeps F90.19 that was made off this loan. That is how it would work for him.

	We credit the Medici for systematizing double-entry bookkeeping, which was defined for them as: \\
		Assets = Liabilities + Equity   \quad \quad \quad \quad \quad \quad \quad \quad \quad \quad \quad \quad \quad \quad \quad \quad (3.1)   \label{eq:Assets}                           
  
Now, note that in this primitive form, the Medici did not have deposits from the borrowers they granted loans to on the books of their banks. In my thought process, what they had when they made a loan like this is a potential liability for the F90.0 they had sent out into the city as paper that was backed by their hoard of metal. Did they record that as a liability? I do not know. In this primitive form, it seems to me that their loan represents a kind of liability. It’s also an asset, because it should return more money than it took from them. 

Only when their borrower is also a client who keeps their accounts at the Medici bank might they see a deposit into the account of the borrower. But, again, nobody is watching over their shoulder. They just have to keep people’s confidence.  

\begin{table}[H]
\begin{adjustwidth}{-1.5in}{0in} % comment out/remove adjustwidth environment if table fits in text column.
\centering
\caption{{\bf Sample loan of 100 florins.} New money is shown in blue. The loan is for F100. F10 is given as gold coin. F90 is given as paper. The debtor takes that out into the world with him and when he comes back he has made a profit over his loan (profit not shown). From that money he pays F100.38 to the bank in gold. Before the reader bounces up and down saying, “But you can’t do that!” think about it. The Medici make a loan that is 90\% paper – all the paper is created money. They get back all of that as hard-money coin, plus hard-money coin as interest. How could Giovanni cancel the hard-money coin away? Looking at the other 90 florin gold coins of principal payment, how would he cancel the latter -- throw it in the Arno River?}
\begin{tabular}{r|rrrrrr}
\textbf{}                      & \textbf{Reserves/Working}       &           &               &                & \textbf{Debtor}         & \textbf{Debtor}     \\
\textbf{}                      & \textbf{ capital}       & \textbf{Liability}           & \textbf{Asset}               & \textbf{Income}               & \textbf{Coin}         & \textbf{paper}     \\
\textit{}                      & \textit{Capital} & \textit{Bank } & \textit{ }    & \textit{}         & \textit{}                    & \textit{}                 \\
\textit{}                      & \textit{account} & \textit{borrowed} & \textit{Suspense}    & \textit{Loan}         & \textit{}                    & \textit{}                 \\
\textit{}                      & \textit{(Tier 1 \& 2 )} & \textit{funds} & \textit{Account}    & \textit{payment}         & \textit{}                    & \textit{}                 \\ \hline
Initial Conditions              & 100                                     &                              &                              &                               & 10                           &                           \\
Loan origination               & -10                                     & {\color[HTML]{CC0000} }      & {\color[HTML]{CC0000} 100}   & {\color[HTML]{CC0000} }       & {\color[HTML]{800000} 10}    & {\color[HTML]{0000FF} 90} \\
Interest                       &                                         & {\color[HTML]{CC0000} }      & {\color[HTML]{CC0000} 0.38}  & {\color[HTML]{CC0000} }       & {\color[HTML]{CC0000} }      & {\color[HTML]{0000FF} }   \\
Interest Payment \#1. & {\color[HTML]{000099} 0.38}             & {\color[HTML]{CC0000} }      & {\color[HTML]{CC0000} -0.38} & {\color[HTML]{0000FF} 0.38}   & {\color[HTML]{0000FF} -0.38} &                           \\
Loan payoff        & {\color[HTML]{0000FF} 100}              & {\color[HTML]{CC0000} }      & {\color[HTML]{CC0000} -100}  & {\color[HTML]{000099} 100}    & {\color[HTML]{000099} -100}  &                           \\
Totals                         & 190.38                                  &                              & 0                            & {\color[HTML]{000099} 100.38} & {\color[HTML]{000099} 9.62}  & 0                        
\end{tabular}
\label{tab6}
\end{adjustwidth}
\end{table}

In the beginning, the Medici had deposits of precious metal, which they managed by using paper instruments to loan it out, and they had their letters of exchange on their books. This could have gotten more complicated as they loaned combinations of paper notes, letters of credit, and physical gold.  

So, given that, what do we plug into the above formula \ref{eq:Assets}? Giovanni has no liabilities here in the beginning, except, arguably, the possibility that the borrower does not pay back his loan. If the borrower does not pay it back, and nobody ever comes to his bank for actual gold, then, Giovanni is out a maximum of F0.19, which is the interest that he owes to his depositor of gold, plus F10.0 which is the amount of gold he gave to the borrower. Since in that time, loans were mostly done secretly, and there was nobody overseeing his bank’s operations except the partners, I would expect that in reality, if Giovanni did not tell anyone about the loan that went bad, that to the extent it was “just paper” he could give his loan a haircut, simply resetting the amount he had loaned to what he was able to collect, or only counting the hard-money currency that he lost. 

Why not? It was all sinful anyway. Usury was a sin, and Cosimo bought himself into heaven with a papal bull he purchased from the Pope. 

Technically speaking, the Medici’s had the liability of the borrower coming into their bank and demanding F90.0 in gold for his paper money. But given the kind of guard that the Medici family had? I do not think that was ever likely to happen. Going out of one’s way to upset or offend one’s investors or bankers is just not something that enters one’s head.  Feeling upset with them, or taking their name in vain in private, yes. But acting on it to their faces is something that just isn’t done if you want to stay in business. Given a business environment that operated more like a modern day mafia, complaining to them or about them too much is unlikely. 

So, as soon as the Medici bank makes the loan, they have that loan waiting in suspense to be paid back. It is, indeed, suspenseful to wait to find out if a loan will be paid back or not. 

More importantly, we see here that while the payment does retire the loan, the money created by the initial loan is not destroyed when it comes back as hard-money coin. Instead, that hard-money is added to the equity in the bank. To cancel the principal for a loan when gold coin was received to pay all of it back would require taking the gold to another banker, depositing it, bringing back the paper note, then accounting for that paper note by destroying it so that the Medici’s could never get their pile of gold coins back from the other banker.  I find it hard to believe that anyone thinking this through would believe the Medici would do this. 

Any gold received as payment of principal would be deposited into their vault to join the rest of the gold hoard, along with its interest. The Medici’s real gross profit on this loan is F90.38 for an investment of: F10 in real gold coin, minus whatever expenses to cover the cost of the paper note. 
	
Of course, there is Giovanni’s overhead for running his bank, which is considerable. He must pay his guards and family retainers plenty to keep them loyal, give out gifts to all the right people, throw parties, build new structures, keep up appearances, etc.. Given the scheme outlined, the Medici family would not have trouble doing so.

\section{Cancellation of loan principal – origin and issues separated from a now ancient debt tradition.}

Over time, I believe that someone as sharp as Giovanni, Cosimo, or Lorenzo would understand that the region could run low on gold and silver to pay off loans with. And, for them to refuse to allow payment in the same forms that they hand out probably would not work. That could raise questions that could prove inconvenient, perhaps terminally so. Consequently, I think that these early bankers could encourage payment in hard-money coin, but they probably could not require it. I could be wrong though.

As a result, the Medici would have to accept their own paper florin notes as well as checks and other bills in return for larger amounts. It would make sense for them to mint some percentage of the gold in their vaults for circulation. It would give people the sense that the paper was worth what it said it was. And, since they had tapped into the pipeline of gold and silver coming into the Vatican,  and, since these early bankers would be accumulating gold in the region, then if they circulated this gold, that would make it possible to collect some of it back as principal payments. 

In the Medici era, gold and silver were “money” to most people, even though the wealthy avoided carrying it, much as today the wealthy avoid carrying large amounts of cash.  The world maintained this strange dichotomy, generation after generation, until the end of the gold standard only 50 odd years ago. Even today, many people think that the world’s money is terminally broken because we are off the gold standard. Money in this view is either gold or silver, but going back to before the Medici’s, money was more often paper and checks and account records as long as everyone thought there was enough gold. 
	
Thinking about this, we can wonder what a Giovanni or a Cosimo might be thinking also, if, instead of gold, they got back money transferred as deposit from another vault operator in another city via a letter of credit? In these cases, I could see them decide that it would be best policy if they were to cancel the principal part of the loan. 
	
I think that the erasing of loan created money when it was paid back was enforced only when the principal repayment was paper. The reason was fear of discovery, and fear of the angry mob that could kill them if they were discovered. Bankers like the Medici had to content themselves with accumulating the interest plus whatever part of the principal did come back as hard-money, together with finding ways to make sure that the interest was covered by gold or silver. 

\subsection{At risk investments and loan principal}

Of course, the Medici engaged in investments on their own accounts, setting up factories in textiles and other goods. Seeing how these transactions were really accounted for would be interesting, however, those records do not exist. Even in the modern era, if a bank makes an investment directly into a real enterprise (at risk investment banking) when the enterprise pays off, this equity is what is received for the loan. And this equity replaces the loan balance. In order to pay down the principal, some amount of that equity must be sold off to someone else. But this is only necessary because of convention. For early bankers like the Medici, why would they do that? Those enterprises were “real things” and the valuation realized from them would not likely be destroyed. I do not believe that the Medici necessarily did cancel the principal on such loans. 
 
\subsection{A philosophical matter in banking – Is a loan real money? Or is it both unreal and real?}

Let me chart this with the most basic kind of loan as shown in table \ref{tab7}. I have simplified this example by making a loan that deposits into the originating bank, and the borrower pays one month of simple interest, then pays it off. Here the created money is shown in blue. I am classing the interest payment as a special category (green text) because it is also a kind of creation of the loan. However, this green money is taken from other existing money/deposits. I am also violating the rule of destruction of principal by payments because here we are looking at the scenario of paying off with gold. Gold, in the Medici’s time was, for historical reasons, what money was. So, it joins the capital of the Medici bank.   

\begin{table}[H]
\begin{adjustwidth}{-1.5in}{0in} % comment out/remove adjustwidth environment if table fits in text column.
\centering
\caption{{\bf Return of a loan after 1 month, interest plus balance.}  Green signifies money that is new to the bank, and kept on the books, but was taken from elsewhere. (e.g. zero-sum money) Blue is money that was created by the loan. Red is a loan that has been written. In this case it is assumed that the money returned is paid back as gold coin, but created by a loan note.}
\begin{tabular}{llllll}
\textbf{}                      & \textbf{Reserves/Working}       &            &                &                &            \\
\textbf{}                      & \textbf{capital}       & \textbf{Liability}           & \textbf{Asset}               & \textbf{Income}               & \textbf{Liability}           \\
\textit{}                      & \textit{Capital account} & \textit{Bank borrowed} & \textit{Suspense}    & \textit{Loan}         & \textit{Deposits}            \\
\textit{}                      & \textit{(Tier 1 \& 2 )} & \textit{funds} & \textit{Account}    & \textit{payment}         & \textit{Deposits}            \\
Initial Conditions             & 100                                     &                              &                              &                               & 10                           \\
Loan origination               &                                         & {\color[HTML]{CC0000} }      & {\color[HTML]{CC0000} 100}   & {\color[HTML]{CC0000} }       & {\color[HTML]{0000FF} 100}   \\
Interest                       &                                         & {\color[HTML]{CC0000} }      & {\color[HTML]{CC0000} 0.38}  & {\color[HTML]{CC0000} }       & {\color[HTML]{0000FF} }      \\
Interest Payment \# 1 on loan. & {\color[HTML]{009900} 0.38}             & {\color[HTML]{CC0000} }      & {\color[HTML]{009900} -0.38} & {\color[HTML]{009900} 0.38}   & {\color[HTML]{800000} -0.38} \\
Payment of loan balance        & {\color[HTML]{0000FF} 100}              & {\color[HTML]{CC0000} }      & {\color[HTML]{CC0000} -100}  & {\color[HTML]{000099} 100}    & {\color[HTML]{000099} -100}  \\
Totals                         & 200.38                                  &                              & 0                            & {\color[HTML]{000099} 100.38} & 9.62           
\end{tabular}
\label{tab7}
\end{adjustwidth}
\end{table}

\subsection{Medici loan principal has a quantum character}

What this thought exercise shows is that, much like quantum mechanics, for the Medici, before the payment of the principal is received, there is uncertainty. The banker knows the payment will be “real money” (e.g. gold), virtual money (money on the books), real goods valued in kind (ex. silk or wool, alum, iron, or bronze), or else some fraction of the principal with a limit of zero. Consequently, the principal of the debt is potentially all of the above – at the same time.  Clearly, the loan is used as money during the term of the loan. However, when it is paid to the bank, if it is “real money” and returned as metal or goods, then it would not get canceled. If it is virtual money, it would be canceled.
 
\subsection{At risk investment of a loan directly by a bank – another form of uncancellable monetary value}

Continuing this exercise regarding cancellation or not of loan principal, we can also think about what happens if the loan is given on the bank’s behalf as an “at-risk” investment. The Medici data strongly suggests that they made at-risk loans on their own behalf. In that instance, what is returned by the loan is the equity in the firm owned by the bank. This equity in a real enterprise, like gold, is real, and it cannot disappear. Even if the Medici did cancel out the principal of the loan on their books, elsewhere, in their holdings, that equity value would appear. So, in effect, in such a case the cancellation of the debt did not really happen. 
	
I have to wonder if JP Morgan did something like this when he backed Edison and Tesla as well. JP Morgan was a major (if not the) driving influence behind the establishment of the Federal Reserve. As a banker, he would have seen directly what Tooke described in his 1844 paper, and was probably familiar with Tooke.  So he should have seen that equity in an enterprise, when it became owned by the bank as the at-risk proceeds of a loan, would cause uncancelled principal funds, whether de facto or de jure. 

\section{Investment banking for venture capital}
Based on the preceding sections, I propose that some changes can be made in banking. I want to be clear that I am differentiating venture capital (VC), which normally produces things of real value in new enterprises, from private equity (PE) which generally does not. PE has a well deserved reputation for raiding and destroying value in existing companies all too often. This is particularly problematic in the USA where laws and legal decisions allow PE operators to buy a firm, take out a loan using the pension fund as collateral, wait 2 years, and then distribute the money from the loan as profits. The company is often taken into bankruptcy as a result, destroying the claims of legitimate parties; the bank collects the pension fund as collateral, and the federal government is left holding the bag to pay pensions. 
 
Allowing PE to use this “legalized theft” mechanism together with non-cancellation of principal, would be devastating to economies that do not have legal frameworks to prevent such destruction of firms for short-term profit. The USA is such a nation. What should occur instead, is to make investing in ventures more desirable than investing in destructive PE \footnote{I do propose that PE not be shut down, but continue to operate as it does today --- not as a banking entity but as private, at-risk, direct investments. In its positive form, PE can consolidate and improve operations.}. 
 
Making at-risk investments is what venture capital does today, except venture capital does it using money in their own accounts, typically provided by limited partners (LPs). This is money-lending, not creation of new money. However, JP Morgan used loans from his bank to fund Tesla and Edison --- in other words, venture capital has devolved in our "modern" world. 
 
 That JP Morgan used at-risk loans from his bank for venture capital investments, and today’s venture capital does not, makes our modern world of finance quite primitive compared to that of 120 years ago. This feature of modern day venture capital is quite odd given banking history, harking back to the Assyrians \cite{Gardiner2006EvoCreidtaryControls}, and the time before banking as we know it was invented\footnote{The Assyrians had state approved businesses quite similar to our corporations.}. It does not make sense to me that VCs make their investments this way today, when 120 years ago, JP Morgan did it using a superior method – creating loans from his bank that he then invested in such enterprises. 
 
I propose a thought experiment regarding the Medici/Renaissance bankers, about what they would do if they made a loan to their own business and that business became worth less than the loan amount. Since in the Renaissance loans and deposits were done in secrecy, nobody except the parties involved would even know the loan had occurred. I am virtually certain that the Medici, and probably other medieval bankers, did not always charge such failed loans as losses to their own accounts.  

\section{Is the double entry loan principal cancellation rule outdated in a fiat system?}
One has to ask, in a fiat money system, if this principal cancellation rule is outdated for some situations. It appears, upon reflection, to exist to prevent the bank from building up reserves and profits that are not gold, silver, or real property equity of some kind. I think this rule’s inception is handed down from a time when the world was switching from hard-currency gold and silver coinage to paper/accounting money. This conversion took centuries. Since someone could ask for the gold coin, at least in theory, not having enough coin to maintain the illusion was dangerous, and this is the basis of “reserves” as they evolved in banking. 

Perhaps we  should be asking: When should this rule be abandoned?  If it should not be maintained, then why should we consider abandoning it? 

If this rule of double entry accounting that requires cancellation of principal of loans is reconsidered, to what extent should that occur? There already are exceptions to this rule. Institutions that are too big to fail which make bad loans to foreign governments are allowed to give those loans a haircut rather than call them in default and cover them with their reserves \footnote{In the case of the post-GFC Greek debt, this prevented hedge funds in the USA from profiting off their holding of CDS notes. The EU Central Bank's logic was that if the loan had a haircut, the losses did not exist. I found this rather amusing.}. Either that, or the central bank buys largely worthless paper at face value, and then “retires” them, as was done after the GFC, which is effectively the same thing. 

My thoughts about a guideline for whether it should be allowed or not is to start with allowing non-cancellation for banks making at-risk investments in equities that directly create real utility value or significant improvements to efficiency, because the bank then receives share(s) in a “real thing” for its investment. However, this question goes deeper than that. 

If cancellation of debt-created principal is done away with for some segment of loan or type of bank, what should happen to principal for ordinary banking operations? Should it all stay with the bank? I do not think that would be wise or helpful. Should it go to fund local government? I think that is a good option.  Alternatively, should the allocation of principal funds be a matter for the central bank to decide, as another method of regulating the economy? This latter could work, as today, allowing this third party to take actions that politicians find difficult --- the guidelines should be for the purpose of ensuring positive equity in the non-governmental sector, allocating to support the smooth operation of state, county, and township government, and to encourage investment in real economy value creation \footnote{Here I differentiate the finance economy from everything else.}. 

\section{Rationale for rewriting the rule on cancellation of loan principal}
	Modern Monetary Theory (MMT) proposes to spend as needed to keep the private sector in positive equity. MMT is founded on the fact that in a fiat system taxation is not required to support spending. However, I think there is a serious problem with this in practice because it eliminates the taxation circuit as a necessity for funding government. I think that this is dangerous for politicians to believe. My concerns are contained in a paper, “Is Modern Monetary Theory’s prescription to spend without reference to tax receipts a recipe for tyranny?” \cite{Hanley2021IsModernMonetaryTheorysprescriptiontospendwithoutreferencetotaxreceiptsaninvitationtotyranny} In summary, this paper says that if politicians no longer believe that they must keep tax revenues coming in for the purpose of financing government’s operations, this means they are not inherently motivated by necessity to keep the broader economy going. I think that is the foundation for a Hunger Games kind of world unless politicians are better people than they appear to be. 
 
\section{Proposal 1 --- Commercial lending allots paid principal to local, state, and federal government}
Here I propose an alternative to the MMT prescriptions. The core of this is to create sound mechanisms for sources of permanent money creation. The key to this is to tie it to creation or maintenance of real economy utility value. Bankers are not so likely to lend unless they are in positive equity, and keeping the government sector in negative equity should accomplish that. Today, keeping the non-government sector in positive equity requires federal government spending just as MMT states. 

In it’s simplest concept, I propose that instead of cancellation of loan principal, that cancelled principal accrue to state, county, and city governance to finance their operation. This would be the entry point of this increase in the money supply that would then be spent into the economy. 
	
Making this change would allow private enterprise, plus local and state government to remain the driver for what government can do, and tie it directly to real economy activity. While there are problems with this such as regulatory capture, appearance of monopoly, monopsony, etc., I think this is preferable to the enablement of tyranny combined with the tools that modern technology makes available. We know how to fight the former battles, difficult as they are. The latter battles with tyranny have proven much less tractable and resolution often bloody. 

An alternative that could be considered, is that the central bank could be assigned to decide what the proportions were that went to each type of government. This would maintain the central bank’s role as key stabilizer to the economy, giving it an additional set of tools to work with. And it would maintain the relative freedom from political pressure that allows immediate and effective action to be taken in a crisis. Within the MMT model, there is justification for giving all of such money to state, county, and city governments.  Leaving that task to the central bank would give some independence for such decisions. If this new mechanism were operated by the central banks, it would leave much of the basic financial circuit of the federal government to operate as it does today. 

\subsection{Running the numbers}
\begin{wrapfigure}[15]{l}{95mm}
\includegraphics[width=95mm]{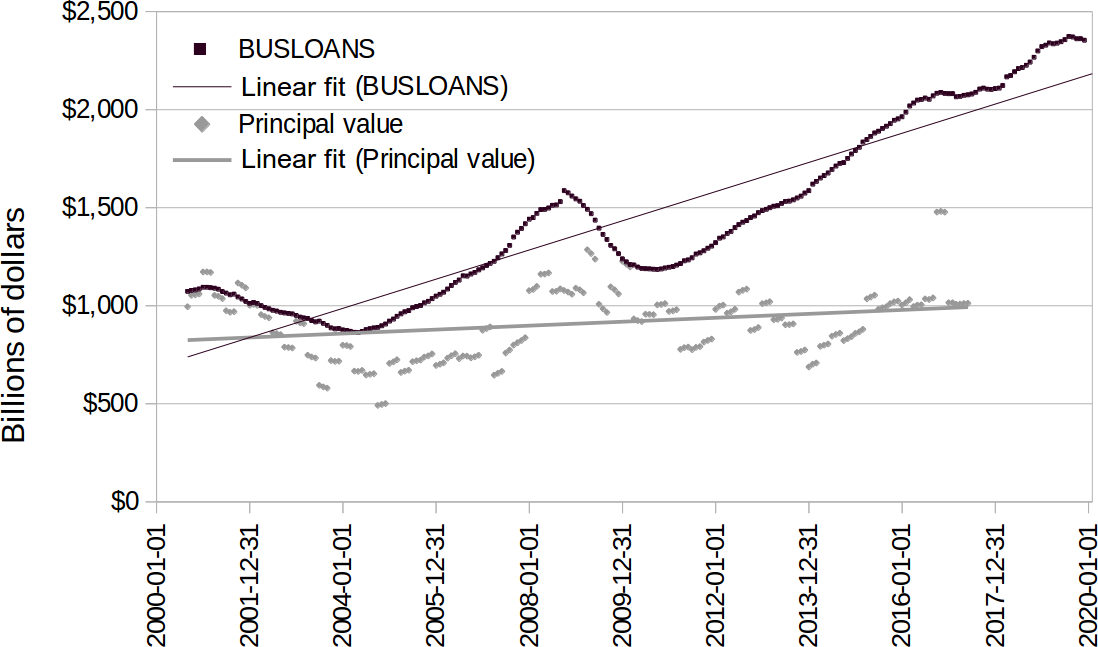}
\captionsetup{labelformat=empty} % makes sure dummy caption is blank
\caption{} 
\label{Fig_5_BUSLOANS_Principal-value} 
\end{wrapfigure}

Reported commercial loan activity (BUSLOANS) in the USA over the years from 2000 to 2019 spans a range from \$862.8 billion to \$2.373 trillion per year with a median of \$1.357 trillion \cite{FRED2020BUSLOANS}.  FRED data for weighted average term of commercial loans (EDANQ) was discontinued in 2017. Normalizing BUSLOANS for the years 2000 to 2017, using EDANQ, generates a range from \$491.4 billion to \$1.481 trillion, with a median of \$926.2 billion \cite{FRED2020EDANQ}. We can see in figure \ref{Fig_5_BUSLOANS_Principal-value} that most of the rise in commercial loan activity is offset by longer terms on the loans. We make a simplifying assumption here that since the terms are so short, principal payments in year 1 and 2 do not change enough to matter. 

\newpage
\marginpar{
\vspace{.7cm} 
\color{Gray} 
\textbf{Figure \ref{Fig_5_BUSLOANS_Principal-value}. FRED:BUSLOANS} 
Commercial loan activity (FRED:BUSLOANS)  vs. normalized value of principal oustanding. 
}
 % avoid blank space here

These data allow us to see in the normalized principal value, approximately what the yield payments could be to state and local government for each year. If this policy were restricted to commercial loan activity only, it could provide up to approximately \$1 trillion or so for local government operations with the current level of commercial activity. And this policy change would incentivize local governments to work to increase business activity. 

State and local governments in the USA collected for 2018 and 2019, \$2.643, and \$2.743 trillion in tax revenue \cite{NIPA2020Table3.20}. If commercial loan principal payments were added to this, for 2018 and 2019, it would add \$1.1 trillion per year, which is an additional 41.6\% and 40.1\% respectively to state and local government. This money would then be maintained in circulation, slowly adding to the positive equity of the non-government sector. Modelling should, of course, be done on this. 

\subsection{Chaining of loans to puff up principal -- a proposal 1 pitfall}
If state and local governments become highly funded by commercial loan activity, this could result in politicians encouraging banks to write N loans, each with an interest rate a fraction of the normal rate such that the final loan is the one paid by the end borrower as shown in equation (1). In this way, the amount of principal could be artificially inflated, eventually breaking the system after a long period of time. So it is preferred that non-cancellation of principal occur for a loan to a direct end-borrower who uses that loan for a business purpose creating real utility value goods or services. 

	Loan 1 ← Loan 2 ←… Loan n ←Borrower \quad \quad \quad \quad \quad \quad \quad \quad \quad \quad \quad \quad (2)

\subsection{Consumer loan payments -- an open question}
	Whether or not consumer loan principal payments should be allocated, in whole or in part, to state and local government in the way that commercial loans are is a good question. On the negative side, the spiking of private, particularly household, debt has been a noted as a problem \cite{Keen2018JapansLessonsForGlobalEconomy}. On the positive side, consumer debt has been used to compensate for lack of government spending and policy that supports the consumer sector. I lean to saying that direct consumer debt principal would help compensate by adding money into the public sector for balance. 

\section{Proposal 2 --- Venture Capital to retain principal and write-down failed loans.}
\subsection{Retention of principal valuation when realized through at-risk equity investment}
	For true venture capital, in which at-risk investments in real enterprises that create real-economy utility value\footnote{i.e. not finance, nor finance of finance, nor purchase of existing business, nor purchase or hypothecation of stocks, FOREX, cryptocurrency, etc.}, the equity valuation of the invested funds should be retained at whatever the book value is, and the principal could be retired without payment. This results in permanent creation of money, and this money is retained in the hands of the investors, generally in the form of stocks, at least initially. The net result should be two desirable outcomes for society. 
	First, this would result in maintenance and slow improvement of the positive equity position of the private sector which keeps the economy ready to invest. Second, this would make investment in ventures that overall benefit society with new value creation to be more profitable for investors than other types of investment. This, in turn, should lower the net cost of investment, and hence lower the necessary profitability of such ventures, with the attendant improvement in job creation as more ventures receive funding. 
	Such a venture capital bank operation should not be allowed to borrow from the discount window of the central bank, under any circumstances. The discount window should be the province of ordinary commercial and retail banking. Likewise, for such a venture capital banking operation, no part of the funds-at-risk used as reserves should be borrowed from an ordinary bank. The current venture capital system in which funds are raised entirely from limited partners should be retained, and those funds become the reserves that determine how much the venture bank can invest. Thus, there would not be settling up on the back end as there is in the commercial/retail banking system. 
\subsection{At-risk equity loan haircuts}
	Using the justification that for the VC bank issuer of an at-risk loan, that VC bank wrote the loan on their own account, and said VC bank only owes the loan to themselves, when a venture capital fund is operated as a this type of bank, it could be allowed to give a haircut to any at-risk equity loan that it made without penalty. This will also lower risk for venture capital that operates within a venture capital banking system \cite{Hanley2019EquityDefaultClawbacksVentureBanking}.
	This will have a side effect of permanent money creation in the commercial and consumer economy. This side effect should act to counter inequality, as venture capital generated spending allocates these created funds broadly, with a great deal going to employment of labor and acquisition of capital goods, in a similar way to government spending.  

\section{Conclusion}
Together, these proposals should make it possible to realize many of the goals of MMT without being completely dependent on the fiat currency issuing government and the vagaries of politicians, and to do so driven by what should be among the most skilled set of investors. These proposals will do so by directing directing investment into utility producing enterprises, and lowering risk of such investment.  A degree of permanent money creation in the economy will begin which in part counters inequality, and over time migrates the total monetary system toward one that operates more compatibly with how economists and the public believe the economy should operate. It should become less destructive to engage in targeting balancing of government books, and help stabilize free society. 

\section*{Appendix --- virtual currency of the Medici era and Babylonians}
\label{Sect:Virtual_currency}
The practice of holding receipts (clay, parchment, paper) for coins (silver or gold usually) over the valuable metal itself is at least 4,000 years old. 

\subsection*{Currencies of the Medici era – Florin, lira-a-fiorino, soldi, \& denari}

There were interesting aspects to the currency of the Medici period that advantaged the wealthy. Gold was reserved for the wealthy, for international trade and for luxuries. Silver was used by everyone else. The exchange rate between silver piccolini and gold florins was based on the exchange rate between the metals, and was not a normal coinage system. This created accounting problems (which were also opportunities for cooking the books) because workers were paid in piccolini, or else in goods. When paid in goods the value of the goods could be played with as well. But factories were valued in, and paid their suppliers and profits in, florins. When first coined, 1 piccolino = 1/20th of a florin. 250 years later, 1 piccolino = 1/140th florin.  

More interesting for the development of currency, were the virtual coins used to keep track of florin accounts. A ‘lira a fiorino’ was 20 29ths (20/29) of a florin. Each of these lira was worth 20 soldi, and each soldi was worth 12 denari. The denari comes from roman currency. This meant that one florin = 29 soldi or 348 denari. 

If you are a numbers person, 29 is the “proper” number of days in a month in the Roman calendar. This corresponds to the simplistic observation of the length of the lunar month which we have now measured at 29.5306 days. But even if the correct value is used for the lunar month, it does not fit correctly into a year.  The number 12 for the denari is the number of months in the year, and the number of signs in the zodiac. 12 is still used today in timekeeping. The month/year is probably the reason for the number 29 appearing in the Florentine currency when converting into this coinage rooted in roman history. One wonders also, if the complexity of this money helped lenders and accountants to embezzle from the rich. 

In any case, none of these coins of the wealthy except the florin ever actually existed physically, and much, if not most, of the florin transactions were purely bookkeeping entries. That meant that wherever there was something “past the decimal point” it was only on the books, and could never be demanded as real coin. This probably was not a large amount of money, as loans would be for F100 to F1000. But there is a deeper implication. 
 
What this virtual coinage indicates is that in terms of florins, for trade, most accounting, including spending of florins by the rich, must have taken place in bookkeeping, rather than with real, physical coin. For the wealthy, this meant that nobody could debase their coins by clipping, or alloying with baser metals. Nor could their money be easily stolen. That was something that happened to ordinary people who used silver piccolini. 

For the bankers, particularly considering Gardiner’s point below regarding the agio and shubati, this would explain how the bankers could take the risk of making loans of money they did not actually have in their vaults.  They were confident that none of their clients would want to take their money out, just come in to see it in the vault from time to time. This could also allow them to absorb losses better. 

The Medici used roman numerals to record most transactions. While one might think this would lead to errors, addition and subtraction with roman numerals is actually easier \cite{DeRoover1966RiseDeclineMedicidBank}. And, the wealthy took checks written out on their bank – another perk of wealth. The origin of this was probably that it made it difficult to rob the wealthy. In a time when robber gangs and mercenaries were rampant, literally causing wealthy castles to have arrangements to pour boiling oil down on brigands, carrying physical cash was unwise. 

\subsection*{The Agio and preference for accounted money since ancient times}
\label{Sect:Agio_shubati}
The agio is the cost charged to free the depositor or merchant from the need to carry gold or silver around. Accounting of agio fees charged by these Italian bankers are not available, but it is known that agio was significant. This term is defined in modern times as simply the money exchange fee.  Gardiner discusses the meaning of agio in the Medici era. 

\begin{quote}
“It would never occur to modern economics students that people preferred a banker's paper to coins and that they were prepared to pay heavy charges [agio] for it. A bill of exchange is as reliable as the debtor who accepted it. There is no question of the soundness of the money. It cannot be debased or clipped.  It makes sense that Babylonians preferred to hold the receipts for silver, 'shubati', of the local temple to holding actual silver. The oldest shubati are 4,000 years old. The British Museum has 650 of them.” - Gardiner, G \cite{Gardiner2019Agio} \end{quote}

\section*{Acknowledgments}
I want to thank Geoffrey Gardiner for his interest and patience in this topic sustained over years. I also want to thank Steve Keen and Russell K. Standish for their years of work on Minsky \cite{Keen2020Minsky}. Minsky has been most helpful although it is not directly cited. 

%\nolinenumbers

%This is where your bibliography is generated. Make sure that your .bib file is actually called library.bib
\bibliography{CancellationOfPrincipal}

%This defines the bibliographies style. Search online for a list of available styles.
\bibliographystyle{abbrv}

\end{document}